\documentstyle[aps,preprint]{revtex}
\begin{document}
\draft


\title{
Numerical transfer-matrix study of metastability in the $d\!=\!2$ Ising model
}

\author{C.~C.~A.\ G{\"u}nther*\dag, Per Arne Rikvold*\dag\ddag,
and M.~A.~Novotny*}
\address{
* Supercomputer Computations Research Institute, Florida State
University, Tallahassee, Florida 32306-4052 \\
\dag Department of Physics and Center for Materials Research and Technology,
Florida State University, Tallahassee, Florida 32306-3016 \\
\ddag Faculty of Integrated Human Studies, Kyoto University, Kyoto 606,
Japan\S
}

\date{\today}
\maketitle

\begin{abstract}
We apply a generalized numerical transfer-matrix method to the
two-dimensional Ising ferromagnet in a nonzero field to obtain
complex constrained free energies. Below $T_c$ certain eigenstates
of the transfer matrix are identified as representing a metastable
phase. The imaginary parts of the metastable constrained free energies
are found to agree with a field-theoretical droplet model
for a wide range of fields, allowing us to numerically estimate the average
free-energy cost of a critical cluster. We find excellent agreement
with the equilibrium cluster free energy obtained by
a Wulff construction with the exact, anisotropic zero-field surface
tension, and we present strong evidence for
Goldstone modes on the critical cluster surface.
Our results are also fully consistent with average
metastable lifetimes from previous Monte Carlo simulations.
The study indicates that our constrained-transfer-matrix technique
provides a nonperturbative numerical method to obtain an analytic
continuation of the free energy around the essential
singularity at the first-order transition.
\end{abstract}
\vspace{0.5 cm}
\pacs{1993 PACS~Numbers: 64.60.My, 64.60.Qb, 82.60.Nh, 05.50.+q}
\narrowtext



Despite the fact that metastability is commonly observed in systems
ranging from fluids, magnets, and alloys, to the ``false vacuum''
in high-energy physics, a fully satisfactory microscopic description
remains elusive. Although metastable states do not globally
minimize the free energy, they persist for periods
many orders of magnitude longer than other characteristic timescales
of the system \cite{pen71,sew80}. This observation has
prompted efforts to study the statistical mechanics of
metastable states as if they were equilibrium phases, using
constrained partition functions that exclude or severely reduce the
probabilities of microstates that dominate in equilibrium \cite{pen71}.
By applying such ideas to a field-theoretical droplet model with Fokker-Planck
dynamics, it has been shown \cite{lan67,lan69}
that the nucleation rate for droplets of the equilibrium phase
in a metastable background is proportional to the imaginary
part of a complex-valued constrained free energy obtained by analytic
continuation from the equilibrium phase into the metastable phase.
This result was derived for ultra-weak fields only \cite{lan69},
and despite extensive subsequent work \cite{new77,mcc78,roe84} its domain
of validity is not completely clear.

Recently one of us introduced a numerical
technique, the Constrained-Transfer-Matrix (CTM) method, which extends
transfer-matrix calculations to metastable and unstable
states \cite{par89}, providing a means to compute complex-valued
``constrained free energies''.
However, in contrast to the field-theoretical droplet model
\cite{lan67,lan69}, the CTM method does not necessitate the
explicit introduction of droplets.
In this Letter we apply
this method to the two-dimensional square-lattice nearest-neighbor
Ising ferromagnet in a magnetic field and below its critical temperature.
Metastable states in short-range force systems have finite,
albeit very long, lifetimes \cite{mcc78,sto72,bin73,sto77,tom92},
and it is now quite well established that the free energy has an
essential singularity at $H$$=$$0$ \cite{isa84,abr92}.
This contrasts sharply with the infinitely long-lived metastability
displayed by systems with weak long-range interactions in the
thermodynamic limit \cite{pen71,sew80,roe84,par89,par92,gor93}
and the corresponding spinodal branch-point singularity.
The numerical results for the metastable free energy
that we present here are in excellent agreement with series
expansions \cite{low80,wal80,bak80,har84}
and Monte Carlo simulations \cite{sto77,tom92,jac83,par93},
and indicate that the CTM technique provides a nonperturbative
numerical method to obtain an analytic continuation of the free
energy around the essential singularity at $H$$=$$0$.

The Ising model is defined by the Hamiltonian ${\cal H} \! = \!
-J \sum_{\langle i,j \rangle} \sigma_i \sigma_j - H \sum_i \sigma_i$,
where $\sigma_i \! = \! \pm 1$ is the spin at site $i$, $J$$>$0 is
the nearest-neighbor coupling, the first sum is over all
nearest-neighbor pairs on the square lattice, and $H$ is the field.

The conventional transfer-matrix formalism \cite{dom60} provides a
method to calculate the equilibrium partition function and
correlation lengths for an $N$$\times$$\infty$ system
from the eigenvalues of the transfer matrix ${\bf T}_0$. This matrix
is defined by dividing the system into one-dimensional layers
of $N$ spins each and by decomposing the total Hamiltonian
(in this case $\cal H$ from above) into partial
Hamiltonians ${\cal H}(X_l,Y_{l+1})$, giving $\langle X_l
|{\bf T}_0| Y_{l+1} \rangle=\exp[-\frac{1}{T} {\cal H}(X_l,Y_{l+1})]$.
Here $T$ is the temperature in energy units, and
$X_l$ and $Y_{l+1}$ are the configurations of the $l$-th and
$(l\!+\!1)$-th layer, respectively.
Since ${\bf T}_0$ is a positive matrix, its dominant eigenvalue
is positive and nondegenerate ({\it i.e.}, $\lambda_0 \! > \! |\lambda_\alpha|$
for all $\alpha \! > \! 0$), and the corresponding eigenvector
$| 0 \rangle$ is the only one whose elements are all positive \cite{dom60}.
The CTM method extends the scope of transfer-matrix calculations
to metastable and unstable states
by defining constrained probability densities constructed
from the eigenvectors $|\alpha \rangle$ of ${\bf T}_0$,
\begin{eqnarray}
P_{\alpha}(X_l,Y_m) & = & \langle \alpha|X_l \rangle \langle X_l
                     |(\lambda_{\alpha}^{-1} {\bf T}_{\alpha})^{|m-l|}
                     |Y_m \rangle \langle Y_m|\alpha \rangle
                                                \nonumber \\
P_{\alpha}(X_l) & = & \langle \alpha| X_l \rangle \langle X_l
                               |\alpha \rangle \;\;.
                                            \label{eq1}
\end{eqnarray}
The matrix ${\bf T}_\alpha$
is chosen to commute with ${\bf T}_0$, but has $\lambda_\alpha$ as its
dominant eigenvalue, rather than $\lambda_0$. This ensures that
$\lim_{|m\!-\!l| \rightarrow \infty}P_\alpha(X_l,Y_m) \! = \! P_\alpha(X_l)
P_\alpha(Y_m)$,
corresponding to stochastic independence of widely separated layers.
As a consequence, the ${\bf T}_\alpha$ for $\alpha$$>$$0$
are {\it not} positive matrices.
As in Refs.~\cite{par89,par92,gor93} we use
\begin{equation}
{\bf T_\alpha} = \lambda_\alpha \left\{
\sum_{\beta < \alpha}
| \beta \rangle {\lambda_\alpha \over \lambda_\beta} \langle \beta |
+ | \alpha \rangle \langle \alpha |
+ \sum_{\beta > \alpha}
| \beta \rangle {\lambda_\beta \over \lambda_\alpha} \langle \beta |
\right\}.
                          \label{eq2}
\end{equation}
{}From ${\bf T}_\alpha$ one obtains constrained free-energy densities,
\begin{mathletters}
\begin{equation}
F_\alpha = {1 \over N} \sum_{X_l,Y_{l+1}}
P_\alpha (X_l,Y_{l+1}){\cal H}(X_l,Y_{l+1}) - TS_\alpha \;\;,
                           \label{eq3}
\end{equation}
where
\begin{equation}
S_\alpha = -{1 \over N} {\sum_{X_l,Y_{l+1}}} P_\alpha (X_l,Y_{l+1})
{\rm Ln} \langle X_l | \lambda_\alpha^{-1}{\bf T_\alpha} | Y_{l+1} \rangle
                           \label{eq4}
\end {equation}
\end{mathletters}
are constrained entropy densities, defined in analogy with the source
entropy of a stationary Markov information source \cite{bla87}.
Negative elements of ${\bf T}_\alpha$ give
rise to the imaginary part of $F_\alpha$ through Ln$z$, the principal
branch of the complex logarithm.
The quantities defined in Eqs.~(\ref{eq2})--(\ref{eq4}) reduce to
their standard equilibrium values \cite{dom60} for $\alpha$=0.

To calculate $F_\alpha$ from Eq.~(\ref{eq3}) one has to
sum over {\it all} the eigenstates of ${\bf T}_0$. To
accomplish this with sufficient accuracy for large $N$,
diagonalization routines in 128-bit precision for the Cray Y-MP/432
vector-supercomputer, involving block-diagonalization of ${\bf T}_0$,
were custom-written. For $T/J \! \leq \! 0.4$,
the attainable values of $N$ were limited by numerical underflow for small
$|H|$, whereas for higher $T$ they were limited by the available memory.
The total computer time spent on this study was on the order of 800 CPU hours.

Following previous transfer-matrix studies \cite{new77,mcc78},
the metastable state for any particular $H$$>$0
was identified as the eigenstate $| \alpha \rangle$ whose
magnetization is closest to $-1.0$. Fig.~1 shows ${\rm Im}F_{\alpha}$ vs.\
$J/H$ for those $\alpha$ that contribute to the metastable
state for $N$=9 and~10. Each of the lobes corresponds to a
different $\alpha$. Their minima, especially for small $H$, have
extremely small values. The crossings correspond to avoided
crossings in the eigenvalue spectrum of ${\bf T}_0$. At
$T$=0 the exact values of $H$ at which these crossings
occur are $H$=0 and $H/J \! = \! 2/(N\!-\!m)$ with
$m \! = \! 1, \ldots ,N\!-\!1$.
For the temperatures studied, the field values
of the crossings deviate by at most 3\% from their values at $T$=0.
As pointed out previously \cite{new77,mcc78}, these crossings
should correspond to zeros of the constrained metastable partition
function, and as $N$$\rightarrow$$\infty$ the diverging density
of crossings near $H$$=$$0$ gives rise to an essential
singularity \cite{abr92}.

We find two different scaling regimes for the
minima of the metastable Im$F_\alpha$.
The minimum which lies in the field interval
$0 \! \leq \! H/J \! \leq \! 2/(N\!-\!1)$
decays exponentially with $N$. The value of $H$ at which that minimum
occurs, is proportional to $N^{-2}$, whereas
the corresponding dominant length scale is proportional to $N$.
For $H/J$ between approximately $2/(N\!-\!2)$ and 2,
the metastable Im$F_\alpha$,
while still extremely small, are independent of $N$.
These results can be explained by a droplet model in which the
initial step in the decay of the
metastable state is the random formation of clusters that exceed a
critical size.

In weak fields, the average critical cluster is larger
than $N$. Therefore, the free-energy cost
of creating a critical cluster is simply the free energy of an interface
cutting across the system, and thus proportional to $N$. Assuming that the
Boltzmann factor of this free-energy cost, as well as the minimum of
the metastable ${\rm Im}F_{\alpha}$,
is inversely proportional to the nucleation rate, one
finds that  ${\rm Im}F_\alpha$ decays exponentially with $N$.
The $N^{-2}$ dependence of the field at which this minimum occurs can also
be derived by a droplet-model argument \cite{ccag93}.

Field-theoretical droplet-model
calculations for infinitely large systems in the limit
of very weak magnetic fields \cite{lan67,guen80} show that
in two dimensions the imaginary part of the metastable free energy
should be given as \cite{har84}
\begin{equation}
{\rm Im}F_{\rm ms} = B |H|^b \exp \left(-\frac{1}{T}\,
                           \frac{\widehat{\Sigma}^2}{4 |H| \Delta M}\left[1
                             + O(H^2)\right] \right),
                                               \label{eq5}
\end{equation}
where $B$ is a nonuniversal constant.
The leading term in the exponential is $1/T$ times the nonuniversal
free-energy cost of a critical cluster, with $\widehat{\Sigma}^2$
being the square of its surface free energy divided
by its area, and $\Delta M$ the difference in magnetization between the
equilibrium and metastable states.
The prefactor exponent $b$ is determined by Goldstone
excitations on the critical cluster surface and is
expected to be universal. For $d$$=$$2$, $b$ is expected to be
unity if Goldstone modes are present
\cite{low80,wal80,har84,guen80,zia85} and $-1$ if they are
absent \cite{wal80}. Series expansions \cite{low80,wal80,bak80} and
Monte Carlo simulations \cite{jac83} support Eq.~(\ref{eq5}) with
$b$=1. Furthermore, there is substantial evidence \cite{har84,jac83}
that, in the limit of weak fields, $\widehat{\Sigma}^2$ is given by
its equilibrium value, which may be obtained by combining a Wulff construction
with the exact, anisotropic zero-field surface tension \cite{zia82},
and that $\Delta M$ is given by twice the exact zero-field magnetization
\cite{har84}.

Despite the fact that Eq.~(\ref{eq5}) was derived in the weak-field limit,
Fig.~1 indicates that it also gives the correct leading-order behavior of the
minima of the metastable ${\rm Im}F_{\alpha}$ for
{\it intermediate} fields, $2/(N\!-\!2) \! \leq \! H/J \! \leq \! 2$.
The dashed straight line in Fig.~1 is drawn through the two
minima between $J/H$=3 and~4 for
$N$=10 only. However, it also connects several minima at smaller
values of $J/H$ for both $N$=10 and~9,
and its slope gives a rough estimate for $\widehat{\Sigma}^2 / \Delta M$
in Eq.~(\ref{eq5}). In this field region
the size of the average critical cluster is smaller than the finite
width of the system. Indeed, we find that the average droplet diameters,
as determined from the droplet model,
range from about 8 lattice units for $N$=10 near the cross-over
field $H/J \! \approx \! 2/(10\!-\!2)$, down to 1 near
$H/J \! \approx \! 2$ for all $T$ studied \cite{ccag93}.
Furthermore, for all temperatures and all fields $H/J$$\lesssim$$1.2$,
we find that both the metastable and the
stable single-phase correlation lengths, as determined from the
transfer-matrix level spacings, are smaller than the average droplet
diameters. Except for the largest fields studied, this ensures sharp
cluster surfaces.

In order to determine $\widehat{\Sigma}^2/\Delta M$ and
$b$ in a more systematic way, we performed a three-parameter linear
least-squares fit of the logarithm of the minima of the metastable
${\rm Im}F_{\alpha}$ to the form of Eq.~(\ref{eq5}).
It must be emphasized, however, that since our results for
${\rm Im}F_{\alpha}$ are numerically exact, the least-squares method
was used merely as a tool to estimate the parameters
numerically. We have therefore assigned error bars
such that $\chi^2$ per degree of freedom is approximately unity.

Our results for $\widehat{\Sigma}^2/\Delta M$ are in agreement
with the hypothesis
that $\widehat{\Sigma}^2$ is equal to its equilibrium value obtained by
combining a Wulff construction with the exact, anisotropic
zero-field surface tension \cite{zia82}, and that $\Delta M$ equals twice
the exact zero-field equilibrium magnetization.
We find agreement to within 10\% for all temperatures.
We therefore fix $\widehat{\Sigma}^2/\Delta M$ to its equilibrium value
and obtain $b$ from a {\it two}-parameter linear least squares fit.
For $T/J$$>$$0.4$, the results are consistent with the presence and
inconsistent with the absence of Goldstone modes on the surface of
the critical clusters, with a value of $b$$=$$0.87 \pm 0.19$ at $T/J$$=1.0$.
The limits on the attainable values of $N$ prevent us from probing
sufficiently small fields to obtain reliable estimates of $b$
for $T/J$$=$$0.4$. Also, large finite-size
effects prevent us from attaining sufficiently high temperatures
to verify any effects of the deviation of $\Delta M$ from~2 \cite{har84}.

Fig.~2 shows a comparison between the equilibrium value of
$\widehat{\Sigma}^2/\Delta M$ with its value obtained from a
{\it two}-parameter fit, in which $b$ is fixed to 1.
It strongly supports the idea that the metastable state
decays through the formation of critical clusters. These are on
the average almost square at low temperatures and nearly circular
at higher temperatures.

Our results are also consistent with direct measurements of
metastable lifetimes by Monte Carlo simulation
\cite{mcc78,sto72,bin73,sto77,tom92,par93},
in which three different scaling regimes are found.
The two low-field regimes correspond directly to those
found here for the metastable Im$F_\alpha$.

In summary, the Constrained-Transfer-Matrix method \cite{par89}
applied to the two-dimensional square-lattice Ising ferromagnet
allows us to numerically obtain complex metastable free energies.
Their imaginary parts depend on system size and magnetic
field in a way fully consistent with the interpretation that this
technique provides a nonperturbative numerical method to obtain
an analytic continuation of the free energy from the equilibrium
phase into the metastable state around the essential singularity
at $H$$=$$0$. However, the CTM method does not rely on the explicit
introduction of clusters of the stable phase in the metastable
background. Comparing ${\rm Im}F_{\alpha}$
with field-theoretical droplet-model calculations of the same
quantity \cite{lan67,har84,guen80}, we estimated the average
free-energy cost of a critical cluster over a wide range
of fields and temperatures, finding excellent quantitative agreement
with theoretical predictions \cite{har84,jac83} involving the
surface free energy of {\it equilibrium} cluster shapes \cite{zia82}.
The above comparison also gives strong evidence for Goldstone
modes on the critical cluster surfaces
\cite{low80,wal80,har84,jac83,guen80,zia85}.
A more detailed account of this study will be reported elsewhere
\cite{ccag93}.

We thank R.~K.~P.\ Zia for helpful correspondence, and
H.~Tomita and S.~Miyashita for useful conversations.
Supported in part by Florida State University through the
Supercomputer Computations Research Institute (Department of
Energy Contract No.~DE-FC05-85ER25000), the Center for
Materials Research and Technology,
and through Cray Y-MP supercomputer time,
and also by National Science Foundation Grant No.~DMR-9013107.



\vskip 0.0 true cm
\eject

\begin{figure}
\caption{The imaginary part of the metastable constrained free energy,
Im$F_{\alpha}$, vs.\ $J/H$ for two different system sizes at $T/J$$=$$1.0$
($T/T_c$$\approx$$0.441$). The metastable Im$F_\alpha$, which
is expected to be proportional to the nucleation rate,
varies over twelve orders of magnitude.
There are two different scaling regions. For small $|H|$,
Im$F_\alpha$ vanishes exponentially with $N$, and for
large $H$ it is independent of $N$. The straight dashed
line is drawn through the two minima marked by stars in
the figure for $N$$=$$10$
only, but it also connects minima at smaller values of $J/H$,
both for $N$$=$$10$ and $N$$=$$9$, as expected from Eq.~(\protect\ref{eq5}).}
\label{fi:1}
\end{figure}



\begin{figure}
\caption{Plot of $\widehat{\Sigma}^2/\Delta M$ vs.\ $T/J$, as
obtained from a two-parameter linear least-squares fit of the
logarithm of Im$F_\alpha$ to Eq.~(\protect\ref{eq5}), setting $b$$=$$1$.
The solid line
corresponds to the equilibrium surface free energy calculated using
a Wulff construction with the exact, anisotropic, zero-field Onsager
surface tension $\sigma(T)$ \protect\cite{zia82} divided by twice the exact
zero-field magnetization $M(T)$. The cluster shape given by
$\widehat{\Sigma}^2$ interpolates between a square
at $T$$=$$0$, given by $8\sigma^2/M$, and a circle
at $T$$=$$T_c$, given by $2\pi\sigma^2/M$, both shown for the
whole temperature range as dashed lines. The metastable transfer-matrix
results follow the equilibrium curve closely. The dotted vertical line
marks $T_c$.}
\label{fi:2}
\end{figure}


\end{document}